\documentclass[aps,prl,twocolumn,secnumarabic,notitlepage,superscriptaddress,nofootinbib,tightenlines,floatfix,preprintnumbers]{revtex4-1}

\usepackage{amsmath,amssymb,amsfonts,mathtools}
\usepackage{float}
\usepackage{graphicx}
\usepackage[colorlinks=true,citecolor=blue,urlcolor=blue,linkcolor=blue]{hyperref}
\usepackage{subfiles}
\usepackage{bbold}
\usepackage{slashed}

\newcommand{\nn}{\nonumber}
\renewcommand{\(}{\left(}
\renewcommand{\)}{\right)}
\renewcommand{\[}{\left[}
\renewcommand{\]}{\right]}
\newcommand{\fnot}[1]{\slashed{#1}}

\begin{document}

\title{Quasi Transverse Momentum Dependent Distributions at Next-to-Next-to-Leading order}

\author{Óscar del Río}
\author{Alexey Vladimirov}
\affiliation{Departamento de F\'{i}sica Te\'{o}rica \& IPARCOS, Universidad Complutense de Madrid, E-28040 Madrid,
Spain}

\begin{abstract}
The partons' transverse momentum can be explored with QCD lattice simulations by studying the quasi-transverse-momentum-dependent parton distribution functions (qTMDPDFs), which are factorized in terms of physical TMDPDFs and soft factors in the limit of the large hadron's momentum. We present the next-to-next-to-leading order (NNLO) calculation of the coefficient function for this factorization. Together with already known expressions for anomalous dimensions, this result allows analysis of lattice data at NNLO perturbative accuracy.
\end{abstract}

\preprint{IPARCOS-UCM-23-032}
\maketitle

\textit{Introduction.} Extraction of parton distribution and related quantities from the lattice simulations is a rapidly growing direction of QCD. The central tool for such extractions is the factorization theorems derived in the large hadron's momentum regime. Combining various equal-time operators and hadron states, one is accessing a variety of parton distributions \cite{Lin:2017snn, Constantinou:2020hdm}. In this work, we study the so-called quasi-transverse momentum dependent (qTMD) parton distribution functions (PDFs) \cite{Constantinou:2020hdm, Constantinou:2022yye}. The corresponding matrix element is defined as (we use the notation of Ref.\cite{Rodini:2022wic})
\begin{eqnarray}
\label{def:quasiTMD}
&&\widetilde{\Omega}_{\text{bare}}^{ij}(y) = 
\\\nn && \quad \langle P|\bar{q}^j(y) [y; b+Lv][b+Lv;Lv][Lv;0]q^i(0)|P\rangle,
\end{eqnarray}
where $|P\rangle$ is the (possibly polarized) hadron state with momentum $P$, $q$ is the quark field, $v$ and $y$ are spacelike vectors (with $y^0=v^0=0$), and $b^\mu=y^\mu-v^\mu (vy)/v^2$. The expression $[a,b]$ represents a straight gauge link from point $a$ to point $b$. The operator represents a quark-antiquark pair separated by $y$ and connected by a staple-shaped Wilson line along direction $v^\mu$ and of size $L$. 

In the regime of the large momentum hadron $P$ and the large length of gauge-link staple $L$, the qTMD matrix element (\ref{def:quasiTMD}) can be expressed via the physical TMD distribution through the factorization theorem, see Eq. (\ref{factorization}), derived using various approaches in Refs.\cite{Ebert:2019tvc, Ji:2019sxk, Ji:2019ewn, Vladimirov:2020ofp, Ebert:2020gxr, Ebert:2022fmh, Rodini:2022wic}. A feature of the qTMD factorization theorem is that in addition to the physical transverse-momentum-dependent parton distribution functions (TMDPDF), it contains an extra function $\Psi$ that accumulates the nonperturbative interaction between the parts of the staple gauge link (also called intrinsic soft factor \cite{Ji:2019sxk}). The treatment of this function is slightly different in different approaches (compare, e.g., Refs.\cite{Ji:2019ewn, Ebert:2020gxr, Vladimirov:2020ofp}), but conceptually the factorization theorem is the same in all cases. The main perturbative ingredient is the coefficient function, which is currently known at next-to-leading order (NLO) \cite{Ebert:2018gzl, Vladimirov:2020ofp}. 

The operator (\ref{def:quasiTMD}) is localized in the equal-time plane, and thus the qTMD matrix element can be simulated within lattice QCD. The results of the simulations can be combined such that the functions $\Psi$ cancel. In this way, one determines the nonperturbative Collins-Soper kernel, see Refs.\cite{Ebert:2018gzl, LatticeParton:2020uhz, Schlemmer:2021aij, Li:2021wvl, LPC:2022ibr, Shu:2023cot}. Alternatively, the $\Psi$ function can be determined from the auxiliary procedure \cite{Ji:2019sxk}, and then one accesses the TMDPDF distribution \cite{LPC:2022zci}. The precision of extraction crucially depends on the accuracy of the perturbative input, which is currently limited by the knowledge of the hard coefficient function. Importantly, this coefficient function is universal and is the same for all polarized quasi-TMDPDF of the leading power \cite{Vladimirov:2020ofp, Ebert:2020gxr}, and it is independent of the particularities of the nonperturbative definition of the internal soft factor.

Nowadays, the extractions of TMDPDFs from the data are routinely performed at next-to-next-to-leading order (NNLO) or N$^3$LO order (see, for example, \cite{Bertone:2019nxa, Scimemi:2019cmh, Bacchetta:2022awv}), and recently were pushed to N$^4$LO order \cite{INPREP}. It has been demonstrated \cite{Scimemi:2017etj, Bertone:2019nxa, Bacchetta:2022awv} that (at least) NNLO is required since the NLO is not sufficiently precise to describe the data from the modern experiments. Modern lattice simulations of quasi-transverse-momentum-dependent parton distribution functions (qTMDPDFs) have yet to reach that order of precision. Still, nonetheless, the NNLO contribution is sizable since the typical scale of lattice simulations is about 1-3 GeV. In this paper, we present the expression for the hard coefficient function for the factorization of qTMDPDF at NNLO. Other perturbative ingredients of the factorization theorem (anomalous dimensions) are known at NNLO and higher. Therefore, using the result of this work, one could analyze the lattice data at complete NNLO.

\textit{Factorization theorem.} The factorization theorem connecting qTMDPDFs and physical TMDPDFs is discussed in many articles \cite{Ebert:2019tvc, Ji:2019sxk, Ji:2019ewn, Vladimirov:2020ofp, Ebert:2020gxr, Ebert:2022fmh, Rodini:2022wic}, which we refer to for detailed discussion. Different spinor components of qTMDPDF correlator (\ref{def:quasiTMD}) obey different kinds of factorization theorems \cite{Rodini:2022wki}. The projection to desired components is done by a contraction with appropriate Dirac matrix
\begin{eqnarray}
\widetilde{\Omega}^{[\Gamma]}(y)=\frac{1}{2}\Gamma_{ji}\widetilde{\Omega}^{ij}(y).
\end{eqnarray}
The most interesting cases are the components projected by $\Gamma\in\Gamma_+ =\{\gamma^+,\gamma^+\gamma^5, i\sigma^{\alpha +}\gamma^5\}$, where $n^\mu$ is a lightlike vector $n^2=0$ defined by the decomposition of hadron's momentum $P^\mu=\bar n^\mu P^++n^\mu M^2/2P^+$ ($P^2=M^2$). For definiteness, we fix
\begin{eqnarray}
v^\mu=\frac{n^\mu-\bar n^\mu}{\sqrt{2}},
\end{eqnarray}
with $v^2=-1$. The components projected by $\Gamma\in\Gamma_+$ obey the leading-power factorization theorem, 
\begin{widetext}
\begin{eqnarray}
\label{factorization}
\Omega^{[\Gamma]}(x,b,\mu)&=&\mathbb{C}_{11}(\mathbf{L}_p,a_s(\mu))\Psi(b;\mu,\bar \zeta)\Phi_{11}^{[\Gamma]}(x,b;\mu,\zeta)
+\mathcal{O}\(\frac{M^2}{x^2(vP)^2},\frac{1}{b^2(vP)^2},\frac{b}{L},\frac{1}{ML}\),
\end{eqnarray}
where
\begin{eqnarray}
\Omega^{[\Gamma]}(x,b;\mu)&=&\int_{-\infty}^{\infty} \frac{dy_v}{2\pi} e^{-ix y_v(vP)}
\widetilde{\Omega}^{[\Gamma]}_{q/h}(y,b; \mu),
\\
\Phi_{11;\text{bare}}{[\Gamma]}(x,b;\mu,\zeta)&=&
\int_{-\infty}^{\infty} \frac{d\lambda}{2\pi} e^{-ix \lambda P_+}
\langle P|\bar{q}(\lambda n+b) [\lambda n+b; b+sn\infty ]\frac{\Gamma}{2}[sn\infty;0]q(0)|P\rangle,
\\
\Psi_{\text{bare}}(b;\mu,\bar \zeta)&=&
\langle 0| \frac{\text{Tr}}{N_c}[-\bar n\infty+b,b][b; b+Lv][b+Lv;Lv][Lv;0][0,-\bar n \infty]|0\rangle
\end{eqnarray}
\end{widetext}
with $y^\mu=y_v v^\mu +b^\mu$, 
\begin{eqnarray}\label{Lp}
\mathbf{L}_p=\ln(\mu^2/(2x(vP))^2),    
\end{eqnarray}
and $a_s=g^2/(4\pi)^2$. The function $\Phi_{11}^{[\Gamma]}$ is the physical TMDPDF of twist-two. The direction of the Wilson lines $s$ is defined by the direction of the staple contour $s=\text{sign}(L)$. In this way, different orientations of the staple contour give access to Drell-Yan or semi-inclusive deep-inelastic scattering (SIDIS) definitions of TMDPDFs, which can be used to test their universality \cite{Musch:2010ka, Musch:2011er}. We stress that the factorization limit is rather complicated (see the last term of (\ref{factorization})). Explicitly, it requires $(vP),L\to \infty$ at fixed-finite $x$ and $b$. We also note that at this power accuracy, there is no difference between $v^-P^+$ and $(vP)=P_z$, which is used as the hard scale. 

The expression (\ref{factorization}) is written in terms of renormalized functions. They are related to the bare functions as
\begin{eqnarray}\label{renormalization-factors}
&&\Omega^{[\Gamma]}(x,b,\mu)= Z_W^{-1}(\mu)Z_{J}^{-2}(\mu)\Omega^{[\Gamma]}_{\text{bare}}(x,b),
\\\nn
&&\Phi_{11}^{[\Gamma]}(x,b;\mu,\zeta)=|Z_{U1}(\mu,\zeta)|^{-2} R^{-1}(b)
\Phi_{11,\text{bare}}^{[\Gamma]}(x,b),
\\\nn
&&\Psi(b;\mu,\zeta)=Z_{\Psi1}(\mu,\zeta)^{-2}Z_W^{-1}(\mu) R^{-1}(b)
\Psi_{\text{bare}}(b).
\end{eqnarray}
Here, the factor $R$ renormalizes rapidity divergences (in most parts of schemes it is equal to the $S^{-1/2}$ where $S$ is the TMD soft factor). The factor $Z_J$ is the renormalization of the quark field in the axial gauge. The factors $Z_{U1}$ and $Z_{\Psi1}$ are ultraviolet (UV) renormalizations of the (leading-twist) semicompact operators constituting the TMD distributions \cite{Rodini:2022wki}. Finally, the factor $Z_W$ depends on $b$ and $L$ and accumulates all divergent factors associated with the staple contour, such as power divergences of spacelike links \cite{Dotsenko:1979wb}, cusps divergences at point $b+Lv$ and $Lv$, and other contributions \cite{Shanahan:2019zcq}. Importantly, the same divergences happen in the function $\Psi$, and thus we do deal with them in our computation of the coefficient function.

The structure of divergences cancellation in the qTMD factorization theorem is similar to those in factorization of Drell-Yan or SIDIS. So, the rapidity divergences cancel in-between soft factor contributions, $\Phi_{11}$ and $\Psi$. The rapidity divergences leave no trace on the coefficient function, but introduce the rapidity scales $\zeta$ and $\bar \zeta$ (see details Refs. \cite{Chiu:2012ir, Echevarria:2012js, Vladimirov:2020umg}). The UV renormalization of TMD distributions cancels the infrared (IR) poles of the coefficient function. The cancellation happens only if
\begin{eqnarray}\label{zeta*zeta}
\zeta \bar \zeta=(2 x \mu (vP))^2.
\end{eqnarray}
The UV poles are renormalized by $Z_J$'s and result in the overall scaling of the qTMDPDF operator.

The scaling of functions (\ref{renormalization-factors}) follows from their renormalization properties. For the $\Omega$ we have
\begin{eqnarray}
\frac{d\ln \Omega^{[\Gamma]}(x,b,\mu)}{d\ln\mu^2} 
=2\gamma_J+\gamma_W(b,L),
\end{eqnarray}
where $\gamma_J$ is the anomalous dimension of the heavy-light current, and $\gamma_W$ is the anomalous dimension associated with the renormalization of the staple link. The evolution of TMD distribution is \cite{Aybat:2011zv, Scimemi:2017etj}
\begin{eqnarray}
\frac{d\ln \Phi_{11}^{[\Gamma]}(x,b;\mu,\zeta)}{d\ln\mu^2} 
&=&\frac{\Gamma_{\text{cusp}}\ln\(\frac{\mu^2}{\zeta}\)-\gamma_V}{2},
\\\label{evol:TMD-zeta}
\frac{d\ln \Phi_{11}^{[\Gamma]}(x,b;\mu,\zeta)}{d\ln\zeta} 
&=&-\mathcal{D}(b,\mu),
\end{eqnarray}
where $\Gamma_{\text{cusp}}$ is the anomalous dimension of the cusp of lightlike Wilson lines, and $\mathcal{D}$ is the Collins-Soper kernel. The Collins-Soper kernel is a nonperturbative function, which represents the interaction of light-quarks in the QCD vacuum environment \cite{Vladimirov:2020umg}. Finally, the $\Psi$ function also evolves with the pair of equations
\begin{eqnarray}\nn
\frac{d\ln \Psi(b;\mu,\zeta)}{d\ln\mu^2} 
&=&\frac{\Gamma_{\text{cusp}}}{2}\ln\(\frac{\mu^2}{\zeta}\)+2\gamma_\Psi
 +\gamma_W(b,L),
\\
\frac{d\ln \Psi(b;\mu,\zeta)}{d\ln\zeta} 
&=&-\mathcal{D}(b,\mu),
\end{eqnarray}
where $\gamma_\Psi$ is the anomalous dimension associated with the finite part of the cusp anomalous dimension at the finite angle. The anomalous dimension\footnote{
By definition the anomalous dimension of the $\Psi$ function is 
$$\gamma_\Psi=-\frac{d \ln Z_\Psi}{d\ln \mu^2}.$$
Evaluating it one should take into account that $\zeta\sim \mu^2$, because $\mu$ is the only dimensional scale of the $\Psi$ function. Therefore, $d\ln(\mu^2/\zeta)/d\ln \mu^2=0$. For the detailed discussion see Ref.\cite{Rodini:2022wic}.}
$\gamma_\Psi$ was computed at LO in Ref.\footnote{
Reference\cite{Vladimirov:2020ofp} provides an incorrect expression for LO $\gamma_\Psi$. This mistake appeared due to the mismatch in definitions for the renormalization constant with earlier paper $Z_J\leftrightarrow Z_J^{-1}$. Once corrected, the expression for $\gamma_\Psi$ coincides with the one computed here or in Ref.\cite{Rodini:2022wic}.} \cite{Vladimirov:2020ofp}, and the NLO term was computed in this work. We found that $\gamma_\Psi$ coincides with the anomalous dimension associated with the heavy-quark \cite{Becher:2009kw} ($v^2>0$). This is not accidental, because the UV renormalization is insensitive to the sign of $v^2$, as it is proven in Ref.\cite{Braun:2020ymy}.

The expressions for anomalous dimensions $\Gamma_{\text{cusp}}$, $\gamma_J$, $\gamma_V$ and $\gamma_\Psi$ are well known. At N$^2$LO they can be found e.g. in Refs.\cite{Braun:2020ymy, Echevarria:2016scs, Bruser:2019yjk}. For the reader's convenience, we have collected all explicit expressions in Appendix A (\ref{a1}) - (\ref{a4}). The remaining anomalous dimension $\gamma_W$ and the Collins-Soper kernel are not important in the present work.

\textit{Quasi-TMD distribution.} The qTMD distribution is an artificial construction that reduces to the physical TMD distribution in the asymptotic limit $(vP)\to\infty$ and $L\to \infty$. Currently, there is no standard construction for this function, see discussion in Ref.\cite{Ebert:2022fmh}. Probably, the most popular way \cite{Ebert:2018gzl, Ebert:2019tvc, Ebert:2020gxr, Ji:2019ewn, Rodini:2022wic} is to consider the function
\begin{eqnarray}\label{def:F-atmu}
F^{[\Gamma]}(x,b;\mu)=\frac{ \Omega^{[\Gamma]}(x,b,\mu)}{\Psi(b,\mu,\mu^2)}.
\end{eqnarray}
Using the factorization theorem (\ref{factorization}), evolution equations and condition (\ref{zeta*zeta}), one finds 
\begin{eqnarray}
F^{[\Gamma]}(x,b;\mu)&=&
\(\frac{(2x(vP))^2}{\zeta}\)^{-\mathcal{D}(\mu)}
\\\nn 
&&
\times \mathbb{C}_{11}(\mathbf{L}_p,\mu)\Phi_{11}^{[\Gamma]}(x,b;\mu,\zeta)+\cdots~,
\end{eqnarray}
where dots indicate the power corrections explicitly given in the last term of Eq.(\ref{factorization}). Note, that in this formulation the scaling equation for qTMD function is
\begin{eqnarray}
\frac{d \ln F^{[\Gamma]}(x,b;\mu)}{d \ln \mu^2}=2(\gamma_J-\gamma_\Psi)
+\mathcal{D}(b,\mu).
\end{eqnarray}
The nonperturbative part of the subtraction factor could be different in other constructions, and it does not affect $\mathbb{C}_{11}$.

Generally speaking, the scales $\mu$ and $\zeta$ are independent; therefore, one can define a more general function
\begin{eqnarray}\label{def:F-atmuzeta}
F^{[\Gamma]}(x,b;\mu,\zeta)=\frac{ \Omega^{[\Gamma]}(x,b,\mu)}{\Psi(b,\mu,\zeta)},
\end{eqnarray}
which reduces to (\ref{def:F-atmu}) at $\zeta=\mu^2$. This function satisfies the pair of equations
\begin{eqnarray}
\frac{d \ln F^{[\Gamma]}(x,b;\mu,\zeta)}{d \ln \mu^2}&=&2(\gamma_J-\gamma_\Psi)
-\frac{\Gamma_{\text{cusp}}}{2}\ln\(\frac{\mu^2}{\zeta}\),
\\
\frac{d \ln F^{[\Gamma]}(x,b;\mu,\zeta)}{d \ln \zeta}&=&+\mathcal{D}(b,\mu).
\end{eqnarray}
Note that the evolution with respect to $\zeta$ has an opposite sign in comparison to ordinary TMD evolution (\ref{evol:TMD-zeta}).

\textit{Coefficient function.} The qTMD operator can be written as a product $J_v^\dagger(y)\Gamma J_v(0)$, where the currents are
\begin{eqnarray}\label{current}
J^{i}_v(0)=[s v\infty,0]q^{i}(0).
\end{eqnarray}
Structurally, the current $J_v$ is similar to the renowned heavy-to-light current (see e.g.\cite{Neubert:1993mb}), with $v^2=-1$, and an open spinor index. The separation between currents $y^2\sim b^2$ is large in comparison to the hard scale $(vP)^{-1}$, and thus any exchange of perturbative gluons between currents is power suppressed \cite{Ji:2019ewn, Vladimirov:2020ofp, Ebert:2020gxr, Rodini:2022wic}. This essentially simplifies the problem of computation of $\mathbb{C}_{11}$ and allows us to present it as
\begin{eqnarray}\label{C11=C1^2}
\mathbb{C}_{11}=|C_1|^2,
\end{eqnarray}
where $C_1$ is the coefficient function for the factorization of a current (\ref{current}) into the leading-twist semicompact operators. Because of this structure, the coefficient function is independent of the $\Gamma$ once $\Gamma\in\Gamma_+$. Therefore, it is universal for all eight leading-power components of the qTMDPDF matrix element.

Comparing the definitions (\ref{factorization}), (\ref{renormalization-factors}), and (\ref{C11=C1^2}) we find that
\begin{eqnarray}
C_1(\mathbf{L}_p,a_s(\mu))&=&Z_{J}^{-1} C_{1,\text{bare}}Z_{U1}Z_{\Psi1},
\end{eqnarray}
where we omit scaling arguments on the right-hand side for brevity. Note that the renormalization constant $Z_{U1}$ and $C_{1,\text{bare}}$ are complex valued in such an approach \cite{Rodini:2022wki}. The renormalization constants $Z_{J}$, $Z_{\Psi1}$ and $Z_{U1}$ are known at N$^3$LO \cite{Baikov:2009bg, Braun:2020ymy, Bruser:2019yjk}. For our NNLO computation, we took the expressions from the appendixes and auxiliary files of Ref.\cite{Braun:2020ymy} (for $Z_{J}$) and Refs.\cite{Echevarria:2015byo, Echevarria:2016scs} (for $Z_{U1}$), and reconstructed from known anomalous dimension \cite{Bruser:2019yjk} (for $Z_{\Psi1}$).

\begin{figure}[t]
\centering
\includegraphics[width=0.45\textwidth]{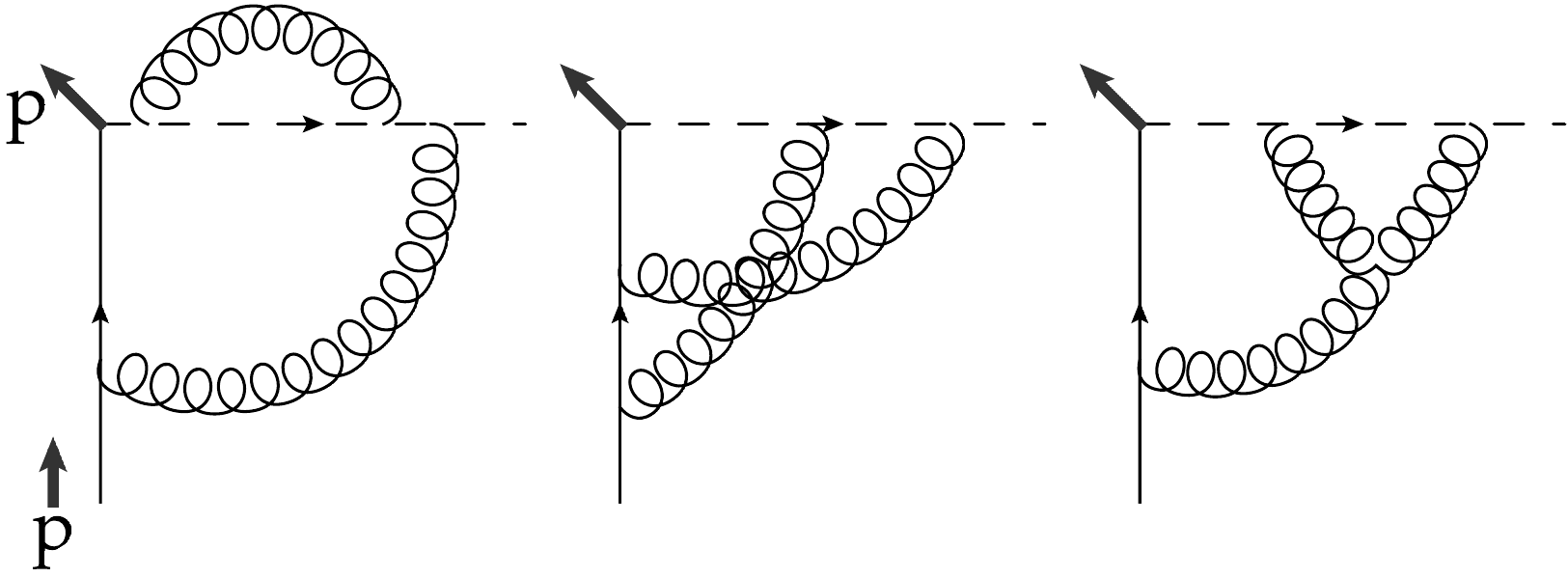}
\caption{Examples of diagrams contributing to $C_1$ at NNLO. The momentum $p$ enters through the quark line and exits from the quark-Wilson line vertex, as indicated on the left diagram.}
\label{fig:diags}
\end{figure}

The examples of diagrams contributing to $C_{1,\text{bare}}$ are shown in Fig.\ref{fig:diags}. Note, that the same diagrams contribute to the computation of the matching coefficient of heavy-light quark current in the heavy-quark effective theory (HQEFT) \cite{Broadhurst:1994se}. The only difference between these computations is the sign of $v^2$ and that the momentum $p$ is passing through the ``light'' quark line, while in the HQEFT matching coefficient computation, the momentum passes through the Wilson line. The computation is done in the dimensional regularization $d=4-2\epsilon$. The reduction to the base integrals is performed by the FIRE6 library \cite{Smirnov:2019qkx}. The result reads
\begin{eqnarray}\label{bareC1}
C_{1\text{bare}}=1+a_s X^\epsilon C_1^{(1)}+a_s^2 X^{2\epsilon} C_1^{(2)}+\mathcal{O}(a_s^3),
\end{eqnarray}
where $X=v^2/(2x(vP)-is0)^2$, 
\begin{eqnarray}\label{C1}
C_1^{(1)}&=&2 C_F\Gamma(-\epsilon)\Gamma(2\epsilon)\frac{1-\epsilon}{1-2\epsilon},
\end{eqnarray}
and the expression for $C_1^{(2)}$ is presented in Appendix A in Eq.(\ref{C2}). Combining $C_{1\text{bare}}$ with the renormalization factors (and renormalizing $a_s$), we observe the exact cancellation of $1/\epsilon$ poles. This provides a general check of the computation.

Substituting the renormalized expression for $C_1$ into Eq.(\ref{C11=C1^2}), we obtain the NNLO coefficient function for the qTMD factorization theorem. It reads
\begin{widetext}
\begin{eqnarray}\label{NNLO}
\mathbb{C}_{11}&=&1+a_sC_F\(-\mathbf{L}_p^2-2\mathbf{L}_p-4+\zeta_2\)+
a_s^2C_F\Bigg\{\frac{C_F}{2}\mathbf{L}_p^4 + \mathbf{L}_p^3
\[2C_F-\frac{11}{9}C_A+\frac{2}{9}N_f\]
\\\nn &&
+\mathbf{L}_p^2\[
C_F\(6-\zeta_2\)+C_A\(-\frac{100}{9}+2\zeta_2\)+\frac{16}{9}N_f\]
+\mathbf{L}_p\Big[C_F\(4+26\zeta_2-24\zeta_3\)
\\\nn && +C_A\(-\frac{950}{27}-\frac{22}{3}\zeta_2+22\zeta_3\) 
+N_f\(\frac{304}{54}+\frac{4}{3}\zeta_2\)\Big]
+C_F\(-12+116\zeta_2-30\zeta_3-\frac{475}{4}\zeta_4\) 
\\\nn &&
+C_A\(-\frac{3884}{81}-\frac{559}{18}\zeta_2+\frac{241}{9}\zeta_3+\frac{99}{2}\zeta_4\)
+N_f\(\frac{656}{81}+\frac{17}{9}\zeta_2+\frac{2}{9}\zeta_3\)\Bigg\}+\mathcal{O}(a_s^3),
\end{eqnarray}
\end{widetext}
where $\mathbf{L}_p$ is defined in Eq.(\ref{Lp}), $C_F=(N^2_c-1)/2N_c$, $C_A=N_c$ are eigenvalues of the Casimir operator of $SU(N_c)$ algebra, $N_f$ is the number of active quarks, and $\zeta_n$ is the Riemann $\zeta$ function. This expression is the main result of this paper. The NLO parts of the coefficient function and anomalous dimension $\gamma_\Psi$ coincide with the known results \cite{Ebert:2018gzl, Vladimirov:2020ofp}.

The logarithm part of the coefficient function can be derived from the evolution equations defined above. It satisfies
\begin{eqnarray}\label{Evolution}
\frac{d\ln \mathbb{C}_{11}}{d \ln \mu^2}=2(\gamma_J-\gamma_\Psi)
+\frac{\gamma_V}{2}-\frac{\Gamma_{\text{cusp}}}{2}\mathbf{L}_p.
\end{eqnarray}
Using explicit expressions for anomalous dimensions (\ref{a1}) - (\ref{a4}) we confirm this. Note that using this equation and (\ref{NNLO}) one is able to compute the logarithm part of the N$^3$LO coefficient function. We present it in Eq.(\ref{NNNLO}).

\begin{figure}[t]
\centering
\includegraphics[width=0.45\textwidth]{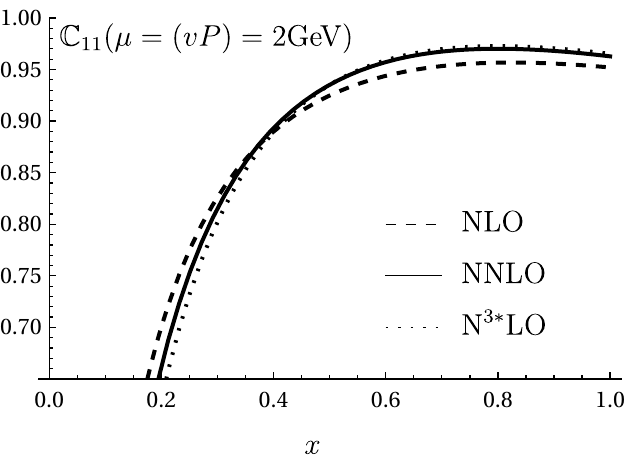}
\caption{Comparison of NLO, NNLO, and N$^{3*}$LO coefficient functions $\mathbb{C}_{11}$ as the function of $x$. The solid, dashed, and dotted lines represent the coefficient functions at NNLO, NLO, and N$^{3*}$LO, correspondingly. The comparison is done for the value of $(vP)=\mu=2$GeV, which is the typical setup for lattice computations.}
\label{fig:plot}
\end{figure}

\textit{Conclusion.} In this work, we have computed the coefficient functions for the factorization of the qTMD matrix at NNLO. We have checked the cancellation of poles between renormalization factors and coefficient function, which provides a check of the factorization theorem for the qTMD operator up to NNLO. These results also allow us to obtain the logarithm part of the N$^3$LO coefficient function. The intermediate and final expressions are also attached to the publication in the \textit{Mathematica} format.

In Fig.\ref{fig:plot} we present the comparison of NLO, NNLO, and N$^{3*}$LO coefficient functions at $(vP)=\mu=2$GeV (i.e. $\mathbf{L}_p=-2\ln x$), where $3*$ indicates that this coefficient function does not have the nonlogarithm term. At these energies, the coefficient function demonstrates a reasonable convergence for $x>0.2$ (NNLO term provides $\sim 5\%$ correction at most). Below $x<0.2$ the convergence drops rapidly; e.g. at $x=0.1$ the NNLO correction is $\sim 20\%$, and N$^{3*}$LO is $\sim 40\%$ (both in comparison to NLO). This shows the natural boundary $x\gtrsim 0.2$ for this approach. To access lower values of $x$ one should find a way to improve the structure of the factorization theorem, either by resumming problematic logarithms (see, discussion on a similar problem for the pseudo-PDF case \cite{Su:2022fiu}) or by matching with a different type of factorization theorem). The coefficient function is just multiplicative, and thus it is straightforward to update the existing procedures including the NNLO correction. It is also important that the coefficient function is independent of the polarization quantum numbers. Therefore, all eight leading-power TMD distributions can be considered at the same NNLO precision.

Let us mention that soon after the publication of the initial version of this manuscript the work \cite{Ji:2023pba} was published. In Ref.\cite{Ji:2023pba} authors derived the same coefficient function at NNLO by studying the threshold logarithms and related functions. Their results agree with ours.

\begin{acknowledgments}
\section*{Acknowledgments}
We thank Ignazio Scimemi, Konstantin Chetyrkin, and Vladimir Braun for helpful discussions. We also thank Yizhuang Liu for spotting a misprint in the initial version of the manuscript, and for helpful comments. AV is funded by the \textit{Atracci\'on de Talento Investigador} program of the Comunidad de Madrid (Spain) No. 2020-T1/TIC-20204. OdR is supported by the MIU (Ministerio de Universidades, Spain) fellowship FPU20/03110. This work is also supported by the Spanish Ministry Grant No. PID2019-106080GB-C21. AV also thanks the DFG FOR 2926 \textit{Mercator Fellowship} for supporting the visit to Regensburg University, where a part of this work has been done.
\end{acknowledgments}

\appendix

\section{Appendix A: Explicit NNLO \& N$^3$LO expressions}

\textit{Bare coefficient function.} The bare coefficient function is defined in Eq. (\ref{bareC1}). Its NLO term is given in Eq.(\ref{C1}) in the closed form. The NNLO term for the bare coefficient function $C_{1\text{bare}}^{(2)}$ has a complicated form involving hypergeometric functions. Here, we present the expression expanded in $\epsilon$. It reads
\begin{widetext}
\begin{eqnarray}\label{C2}
C_1^{(2)}&=& C_F e^{-2\epsilon \gamma_E}\Bigg\{
\frac{C_F}{2\epsilon^4}
+\frac{1}{\epsilon^3}\(C_F-\frac{11}{12}C_A+\frac{N_f}{6}\)
+\frac{1}{\epsilon^2}\[
C_F\frac{5}{2}(1+\zeta_2)
+C_A\(-\frac{133}{36}+\frac{\zeta_2}{2}\) +\frac{11}{18}N_f\]
\\\nn &&
+\frac{1}{\epsilon}\[ C_F\(5+12\zeta_2-\frac{25}{3}\zeta_3\) 
+C_A\(-\frac{673}{54}-\frac{143}{12}\zeta_2+\frac{11}{2}\zeta_3\)
+N_f\(\frac{56}{27}+\frac{13}{6}\zeta_2\)\]
\\\nn &&
+C_F\(4+\frac{145}{2}\zeta_2-\frac{59}{3}\zeta_3-\frac{321}{8}\zeta_4\) 
+C_A\(-\frac{3130}{81}-\frac{2089}{36}\zeta_2+\frac{395}{18}\zeta_3+\frac{159}{4}\zeta_4\)
\\\nn &&
+N_f\(\frac{544}{81}+\frac{143}{18}\zeta_2-\frac{13}{9}\zeta_3\)
+\mathcal{O}(\epsilon)\Bigg\},
\end{eqnarray}    
where we extracted explicitly the $\overline{\text{MS}}$ factor.

\textit{Anomalous dimensions at NNLO.} There are four anomalous dimensions appearing in this problem. Their NNLO expressions are
\begin{eqnarray}\label{a1}
\Gamma_{\text{cusp}}&=&4a_s C_F+4 a_s^2 C_F\[C_A\(\frac{67}{9}-2\zeta_2\)-\frac{10}{9}N_f\]
\\\nn &&
+4a_s^3C_F\Big[C_A^2 \left(\frac{245}{6}-\frac{268 }{9}\zeta_2+\frac{22 }{3}\zeta_3+22 \zeta_4\right)+C_F N_f \left(-\frac{55}{6}+8 \zeta_3\right)
\\\nn &&
+C_A N_f \left(-\frac{209}{27}+\frac{40 }{9}\zeta_2-\frac{28 }{3}\zeta_3\right)-\frac{4}{27}N_f^2\Big]+\mathcal{O}(a_s^4)
\end{eqnarray}
\begin{eqnarray}
\gamma_V&=&
-6a_s C_F+a_s^2 C_F\[C_F
(-3+24\zeta_2-48\zeta_3)+C_A\(-\frac{961}{27}-22\zeta_2+52\zeta_3\) 
+N_f\(\frac{130}{27}+4\zeta_2\)\]
\\\nn &&
+a_s^3C_F\Big[C_F^2 (-36 \zeta_2-136 \zeta_3-288 \zeta_4+64 \zeta_2\zeta_3+480 \zeta_5-29)
\\\nn &&
+C_FC_A \left(\frac{820 }{3}\zeta_2-\frac{1688 }{3}\zeta_3+\frac{988 }{3}\zeta_4-32 \zeta_2\zeta_3-240 \zeta_5-\frac{151}{2}\right)
\\\nn &&
+C_FN_f \left(-\frac{52 }{3}\zeta_2+\frac{512 }{9}\zeta_3-\frac{280 }{3}\zeta_4+\frac{2953}{27}\right)
\\\nn &&
+C_A^2 \left(-\frac{14326 }{81}\zeta_2+\frac{7052 }{9}\zeta_3-166 \zeta_4-\frac{176 }{3}\zeta_2\zeta_3-272 \zeta_5-\frac{139345}{1458}\right)
\\\nn &&
+C_A N_f \left(\frac{5188 }{81}\zeta_2-\frac{1928 }{27}\zeta_3+44 \zeta_4-\frac{17318}{729}\right)+N_f^2 \left(-\frac{40 }{9}\zeta_2-\frac{16 }{27}\zeta_3+\frac{4834}{729}\right)\Big]+\mathcal{O}(a_s^4)
\end{eqnarray}
\begin{eqnarray}
\gamma_J&=&\frac{3}{2}a_s C_F+a_s^2C_F\[C_F\(-\frac{5}{4}+8\zeta_2\)+C_A\(\frac{49}{12}-2\zeta_2\)-\frac{5}{6}N_f\]
\\\nn &&
+a_s^3  C_F \Big[C_F^2 \left(\frac{37}{4}-32 \zeta_2+18 \zeta_3+40 \zeta_4\right)+C_F C_A \left(\frac{655}{72}+\frac{592 }{9}\zeta_2-\frac{71 }{3}\zeta_3+8 \zeta_4\right)
\\\nn &&
+C_F N_f \left(-\frac{235}{18}-\frac{112 }{9}\zeta_2+\frac{44 }{3}\zeta_3\right)+C_A^2 \left(-\frac{1451}{216}-\frac{130 }{9}\zeta_2+\frac{11 }{3}\zeta_3+12 \zeta_4\right)
\\\nn &&
+C_A N_f \left(\frac{128}{27}+\frac{28 }{9}\zeta_2-\frac{38 }{3}\zeta_3\right)-\frac{35}{54}N_f^2\Big]+\mathcal{O}(a_s^4),
\end{eqnarray}
\begin{eqnarray}
\label{a4}
\gamma_\Psi&=&a_sC_F +a_s^2 C_F\Big[C_A\(\frac{49}{9}-2\zeta_2+2\zeta_3\)-\frac{10}{9}N_f\Big]
\\\nn &&
+a_s^3C_F\Big[C_A^2 \left(\frac{343}{18}-\frac{304 }{9}\zeta_2+\frac{370 }{9}\zeta_3+22 \zeta_4+4 \zeta_2\zeta_3-18 \zeta_5\right)+C_F N_f \left(-\frac{55}{6}+8 \zeta_3\right)
\\\nn &&
+C_A N_f \left( -\frac{89}{27}+\frac{40}{9} \zeta_2 -\frac{124}{9} \zeta_3\right) -\frac{4}{27} N_f^2\Big]+\mathcal{O}(a_s^4).
\end{eqnarray}
Here, the expressions for $\Gamma_{\text{cusp}}$ and $\gamma_V$ are taken from Ref.\cite{Echevarria:2015uaa}, the expression for $\gamma_J$ is taken from Ref.\cite{Chetyrkin:2003vi}, and the expression for $\gamma_{\Psi}$ is taken from Ref.\cite{Bruser:2019yjk}. 

Using these expressions and Eq. (\ref{NNLO}), together with Eq.(\ref{Evolution}), we are able to determine the logarithmic part of the N$^3$LO coefficient function 
\begin{eqnarray}\label{NNNLO}
\mathbb{C}_{11}^{(3)}&=&C_F\Bigg\{-\frac{C_F^2}{6}  \mathbf{L}_p^6+\mathbf{L}_p^5\left[-C_F^2+\frac{11}{9}C_F C_A -\frac{2}{9}C_FN_f\right]
\\ \nn &&
+\mathbf{L}_p^4\left[C_F^2\left(-4+\frac{\zeta_2}{2}\right)+C_FC_A \left(\frac{122}{9}-2 \zeta_2\right)-\frac{20}{9}C_FN_f-\frac{121}{54}C_A^2+\frac{22}{27}C_AN_f-\frac{2}{27}N_f^2\right]
\\ \nn &&
+\mathbf{L}_p^3\left[C_F^2\left(-\frac{16}{3}-26 \zeta_2+24 \zeta_3\right) +C_F C_A \left(\frac{1682}{27}+\frac{19 }{9}\zeta_2-22 \zeta_3\right)\right.
\\ \nn &&
\left.+C_F N_f\left(-\frac{254}{27}-\frac{10 }{9}\zeta_2\right)+C_A^2\left(-\frac{2506}{81}+\frac{44 }{9}\zeta_2\right)+C_A N_f\left(\frac{842}{81}-\frac{8 }{9}\zeta_2\right)-\frac{64}{81}N_f^2\right]
\\ \nn &&
+\mathbf{L}_p^2\left[C_F^2\left(12-170 \zeta_2+78 \zeta_3+\frac{475 }{4}\zeta_4\right) +C_FC_A\left(\frac{11996}{81}+\frac{2327}{18}\zeta_2-\frac{1429 }{9}\zeta_3-\frac{89}{2}\zeta_4\right)\right.
\\ \nn &&
\left.+C_F N_f\left(-\frac{2047}{162}-\frac{193 }{9}\zeta_2+\frac{70 }{9}\zeta_3\right)+C_A^2\left(-\frac{29351}{162}+\frac{26}{9}\zeta_2+\frac{220}{3}\zeta_3-22 \zeta_4\right)\right.
\\ \nn &&
\left.+C_A N_f\left(\frac{4469}{81}+\frac{16 }{3}\zeta_2-\frac{16 }{3}\zeta_3\right)+N_f^2\left(-\frac{292}{81}-\frac{8 }{9}\zeta_2\right)\right]
\\ \nn &&
+\mathbf{L}_p\left[C_F^2\left(44-430 \zeta_2+124 \zeta_3+\frac{487 }{2}\zeta_4+8 \zeta_2\zeta_3+240 \zeta_5\right)\right.
\\ \nn &&
\left.+C_F C_A\left(\frac{5704}{81}+\frac{32521}{27}\zeta_2-\frac{6212 }{9}\zeta_3-\frac{2450}{3}\zeta_4+6 \zeta_2\zeta_3-120 \zeta_5\right) \right.
\\ \nn &&
\left.+C_F N_f\left(\frac{6943}{162}-\frac{5374 }{27}\zeta_2+\frac{244 }{3}\zeta_3+\frac{350 }{3}\zeta_4\right)\right.
\\ \nn &&
\left.+C_A^2\left(-\frac{723611}{1458}-\frac{21560}{81} \zeta_2+\frac{13858 }{27}\zeta_3+260 \zeta_4-\frac{112}{3}\zeta_2\zeta_3-100 \zeta_5\right)\right.
\\ \nn &&
\left.+C_A N_f\left(\frac{102683}{729}+\frac{6584}{81}\zeta_2-\frac{608}{9}\zeta_3-44 \zeta_4\right)+N_f^2\left(-\frac{6184}{729}-\frac{128 }{27}\zeta_2-\frac{16 }{27}\zeta_3\right)\right]+
c_3\Bigg\},
\end{eqnarray}
where $c_3$ is the unknown finite part.
\end{widetext}

\section{Appendix B: Evaluation of diagrams}

In this appendix, we provide extra notes about the computation of diagrams for the coefficient function at NNLO. The examples of diagrams are shown in Fig.\ref{fig:diags}. In total, there are 10 diagrams (including self-energy graphs). 

The momentum enters the diagram via the quark line and goes out in the vertex. There is no momentum incoming into the Wilson line.  For example, the second diagram shown in Fig.\ref{fig:diags} reads
\begin{eqnarray}\label{app:example}
I&=&-g^4C_F\(C_F-\frac{C_A}{2}\)\int \frac{d^dk}{(2\pi)^d}\frac{d^dl}{(2\pi)^d}
\\\nn &&\times \frac{(\fnot P+\fnot k)\fnot v (\fnot P+\fnot l)\fnot v u(P)}{[(P+k)^2+i0][l^2+i0][(P+l)^2+i0]}
\\\nn
&&\times\frac{1}{[(k-l)^2+i0][k\cdot v+i0][(k-l)\cdot v+i0]}\,,
\end{eqnarray}
where $d=4-2\epsilon$ is the parameter of dimensional regularization. It is important to keep track of $+i0$ prescriptions because incorrect prescriptions could lead to an improper sign of the resulting integral.

For the leading-power computation, it is sufficient to consider $p^2=0$. Then the only dimensional parameter is $\omega=2(Pv)$. Also, only the good component (with respect to $P$) of the quark field contributes to the leading-power term. To project the corresponding component we do
\begin{equation}\label{Trace}
I^T=\frac{1}{4}\text{Tr}\left[I\gamma^-\gamma^+\right],
\end{equation}
where $I$ is the diagram without a spinor multiplier. After projecting, the diagram decomposes into a sum of simpler scalar integrals, which have the general form
\begin{eqnarray}
&& F(a,b,c,d,e,f,g,h)=\int \frac{d^dk}{(2\pi)^d}\frac{d^dl}{(2\pi)^d}
\\\nn &&\qquad \times \frac{1}{[k^2]^a[(P+k)^2]^b[l^2]^c[(P+l)^2]^d}
\\\nn &&\qquad \times\frac{1}{[(k-l)^2]^e[k\cdot v]^f[l\cdot v]^g[(k-l)\cdot v]^h},
\end{eqnarray}
where all propagators have the $+i0$ pole prescription. Next, the integrals are reduced to the set of base integrals by the integration-by-parts relations (see Refs.\cite{Broadhurst:1994se,Chetyrkin:2003vi} for a similar example). Specifically, we have used the FIRE6 library \cite{Smirnov:2019qkx}. Most parts of the base integrals are evaluated using successively one-loop integrals and are expressed with products of gamma functions. We found only two integrals with nontrivial topology. These integrals can be computed in the terms of higher-order hypergeometric functions or as an $\epsilon$ series \cite{Smirnov:2012gma} (for instance, we have used the Mellin-Barnes method). The results are
\begin{eqnarray}
&&F(0,1,0,1,1,1,0,1)=
\\\nn &&
\qquad
\frac{[v^2-i0]^{-1+2\epsilon}}{[\omega-i0]^{4\epsilon}}e^{-2\gamma_E\epsilon}
\Big(
 -\frac{\zeta_2}{\epsilon}-2\zeta_2-2\zeta_3+\cdots\Big),
\\
&& F(0,1,1,0,1,1,1,0)=
\\\nn &&
\qquad
\frac{[v^2-i0]^{-1+2\epsilon}}{[\omega-i0]^{4\epsilon}}e^{-2\gamma_E\epsilon}
\Big(
 -\frac{\zeta_2}{\epsilon}-6\zeta_2+3\zeta_3+\cdots\Big),
\end{eqnarray}
where $\gamma_E$ is the Euler-Mascheroni constant, and dots represent higher powers of $\epsilon$ series. 

Finally, by collecting the expressions together and expanding gamma functions in $\epsilon$ we obtain the bare expressions for each diagram. For example, the diagram (\ref{app:example}) produces the following expansion:
\begin{equation}
\begin{split}
I=&a_s^2X^{2\epsilon}e^{-2\epsilon \gamma_E}C_F\Big(C_F-\frac{C_A}{2}\Big)\Bigg(\frac{1}{12 \epsilon ^4}+\frac{1}{3 \epsilon ^3}
\\ &
+\frac{1}{\epsilon ^2}\Big(\frac{5}{3}-\frac{3}{4}\zeta_2\Big)
+\frac{1}{\epsilon }\Big(2 \zeta_2-\frac{92}{9}\zeta_3+\frac{47}{6}\Big)
\\ &
+22 \zeta_2-\frac{215}{9}\zeta_3-\frac{1031}{16}\zeta_4+\frac{211}{6}+\cdots\Bigg),
\end{split}
\end{equation}
where $a_s=g^2/(4\pi)^{d/2}$ and $X=v^2/(\omega-i0)^2$.

\bibliographystyle{apsrev4-1}
\bibliography{bibFILE}

\end{document}